\documentclass[aps,pre,showpacs,twocolumn,reprint,floatfix,groupedaddress]{revtex4}
\usepackage{epsfig}

\newsavebox{\astrutbox}
\sbox{\astrutbox}{\rule[-5pt]{0pt}{20pt}}

\def\a{\alpha}

\def\dd{\mbox{d}}
\def\ve{\varepsilon}

\def\s{\sigma}

\newcommand{\gbout}[1]{} 

\begin{document}


\title{Dynamic Boundaries in Asymmetric Exclusion Processes}

\author{Sarah A. Nowak$^1$, Pak-Wing Fok$^{1,2}$, and Tom Chou$^{1,3}$}
\affiliation{$^1$Dept. of Biomathematics, UCLA, Los Angeles CA 90095-1766 \\
$^2$Dept. of Applied \& Computational Math., Caltech, Pasadena CA 91125\\
$^3$Dept. of Mathematics, UCLA, Los Angeles CA 90095-1555}

\date{\today}

\begin{abstract}
We investigate the dynamics of a one-dimensional asymmetric exclusion
process with Langmuir kinetics and a fluctuating wall.  At the left
boundary, particles are injected onto the lattice; from there, the
particles hop to the right. Along the lattice, particles can adsorb or
desorb, and the right boundary is defined by a wall particle.  The
confining wall particle has intrinsic forward and backward hopping, a
net leftward drift, and cannot desorb.  Performing Monte Carlo
simulations and using a moving-frame finite segment approach coupled
to mean field theory, we find the parameter regimes in which the wall
acquires a steady state position.  In other regimes, the wall will
either drift to the left and fall off the lattice at the injection
site, or drift indefinitely to the right.  Our results are discussed
in the context of non-equilibrium phases of the system, fluctuating
boundary layers, and particle densities in the lab frame versus the
frame of the fluctuating wall.
\end{abstract}


\maketitle 
\section{Introduction}

Asymmetric exclusion models with a fixed
\cite{schutz,Derrida-92,Derrida-93}, and typically large number of
lattice sites have been the subject of much recent theoretical
attention \cite{Sandow-94,Kolomeisky-98,Chou-04,Lak-03,
Shaw-03,Evans-94,Lahiri-97,Lahiri-00,Naimi-05,Evans-03,Frey-03,Frey-04}. Biophysical
applications and new fundamental understanding of non-equilibrium
steady-states (NESS) have motivated many extensions of the simple
totally asymmetric exclusion process (TASEP) with open
boundaries. These include partially asymmetric models, where particles
can hop backward \cite{Sandow-94}, exclusion processes with nonuniform
hopping rates \cite{Kolomeisky-98,Chou-04,Lakatos-06}, exclusion among
particles of arbitrary size \cite{Lak-03,Shaw-03,dong}, multispecies
exclusion processes \cite{Evans-94, Lahiri-97,Lahiri-00,Naimi-05},
multichannel exclusion processes \cite{junction} and exclusion
processes with Langmuir type adsorption and desorption kinetics
\cite{Evans-03,Frey-03,Frey-04}. All of these studies have considered
open, well-defined boundaries, where the length of the lattice is
fixed. TASEP models with one open and one closed boundary conditions
have also been considered \cite{Klumpp-05}.

However, applications may arise where the length of the system is
dynamically varying.  The system size may vary because a single
particle pushes against a boundary-defining wall. One example is
helicase-induced opening of replication forks in DNA processing
\cite{Betterton-03}.  Here, the moving replication fork defines a
moving boundary of the system.  Examples of variable-system size
exclusion processes that involve multiple motor particles include mRNA
translation in the presence of hairpins in the mRNA, and molecular
motors processing on elongating actin filaments.  Ribosomes that
process along mRNA (in the process of protein synthesis) during
translation \cite{MacDonald-68,Chou-04} often encounter a hairpin and
the position at which the hairpin starts represents a wall over which
the processing ribosomes cannot pass.  The detachment rates of the
ribosomes and the tightness of the hairpin may determine if the
ribosomes can translate the mRNA through the hairpin sequences.  Actin
polymerization at the leading edge of filopodia also seems to be
mediated by processing molecular motors that may carry actin assembly
components \cite{mitchinson,traffic}. The motors detach, and possibly
attach, anywhere along the growing actin filament
\cite{Klumpp-04}. The depolymerization of the leading tip may be
limited or enhanced by the presence of a motor or other actin
associated proteins \cite{purich,julicher}. Finally, a recent model of
a dynamically extending exclusion process without Langmiur kinetics
has been recently studied \cite{HYPHAE2}.  This model has been applied
to filamentous hyphae growth in fungi \cite{HYPHAE}.

\begin{figure}[h]
\begin{center}
\includegraphics[height=0.48in]{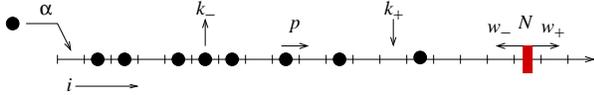}
\end{center}
\vspace{-4mm}
\caption{A totally asymmetric exclusion process bounded by a
fluctuating wall.  Particles are injected onto the leftmost site with
rate $\alpha$, and move to the right with rate $p$.  In the interior,
particles detach and adsorb with rates $k_{-}$ and $k_{+}$,
respectively where $k_{\pm} \ll p$.  The lattice is bounded on the
right by a fluctuating wall with intrinsic hopping rates $w_+$ and
$w_-$, where $w_+<w_-$.}
\label{FIG1}
\end{figure}

With the above applications in mind, we consider a TASEP with a
dynamically varying length.  Specifically, we analyze a many-particle
asymmetric exclusion process with a fixed open boundary on the left, a
fluctuating boundary on the right, and Langmuir kinetics. The
particles have a fixed injection site and can adsorb and desorb. A
wall with an intrinsic leftward drift (representing {\it e.g.}, a
hairpin which energetically favors spontaneous closing or the barbed
end of an actin filament that prefers depolymerization) prevents the
passage of particles. The particles advance and provide a pressure
against the wall. For certain attachment/detachment and wall hopping
rates, the system reaches a NESS in which the statistics of the wall
position are stationary. For other values of the kinetic parameters,
no time-independent  mean wall position exists. The wall will either
drift steadily towards the particle injection site and fall off the
lattice, or move indefinitely away from the injection site,
continuously increasing the size of the system.  The specific details
of the stochastic process are shown in Figure \ref{FIG1}.  Particles
are injected into the first lattice site with rate $\a$ provided it is
empty. In the interior of the lattice, each particle moves forward
with rate $p$ only if the site ahead of it is unoccupied. Particle
attachment and detachment occur with rate $k_{+}$ and $k_{-}$,
respectively, throughout the lattice.

The lattice length is not fixed, and $N$ denotes the position of the
particle-confining wall that hops forward with rate $w_+$, and
backward with rate $w_-$ provided there is no particle to its
immediate left.  The particle occupation at each site, $1\leq i \leq
N-1$, left of the wall is represented by the occupation variable
$\s_{i} \in \{0,1\}$.  If $w_{-} \leq w_{+}$, the wall will move
indefinitely away from the injection site.  In order to prevent the
wall from always escaping to infinity, we consider the more
interesting case of an intrinsic leftward drift described by $w_{-}>
w_{+}$.

The wall position $N$ is not fixed (even at steady state), but rather,
is determined by the intrinsic wall hopping rates, and the
exclusionary interactions between the wall and the lattice
particles. Our analysis is aimed at understanding how the wall
dynamics depend on the parameters $\alpha, k_{\pm}, w_{\pm}, p$. In
the next section, we derive relations for the distribution functions
of the wall.  In steady-state, these relations constrain the particle
density at the wall.  In Section \ref{SIGMA}, we use mean field theory
(MFT) to solve for the density profile, and show that the density
profile obtained using mean field theory is inaccurate near the wall.
Since quantitative prediction of the wall dynamics will require
accurate determination of the particle densities near the wall, in
Section \ref{sec:FS}, we develop a moving-frame finite segment mean
field approach to accurately solve for the density profile near the
wall.  The existence of a steady-state solution and the dependencies
of the mean wall position $\langle N \rangle$ on the problem
parameters are explored and plotted in Section \ref{RESULTS}.

\section{Wall dynamics}\label{WALL}
The net drift of the wall is the difference between its forward and
effective backward hopping rates.  The effective backward hopping rate
depends on both the intrinsic backward hopping rate $w_-$, and on the
occupancy of the site immediately to the left of the wall since a
particle there will block the wall's backward motion.  The wall's
rightward hopping is never impeded.  The probability of finding a
particle immediately to the wall's left varies with its position,
thus, the wall dynamics are position-dependent.  Define $Q_N(t)$ as
the probability that the wall is at position $N$ at time $t$, and
$Q'_N(t)$ as the joint probability that the wall is at position $N$ at
time $t$ \emph{and} the site just before the wall is empty. The wall
dynamics obey

\begin{equation}
\dot{Q}_{N}(t) = w_{-}Q'_{N+1}-w_{-}Q'_{N}-w_{+}Q_{N} + w_{+}Q_{N-1},
\label{EFFECTIVEQ}
\end{equation}
and the moments of the wall position can be formally expressed as
\begin{equation}
{\partial \over \partial t} \langle N^k\rangle = 
\sum_{N=0}^{\infty}(w_{-}Q_{N}'-w_{+}Q_{N})\sum_{j=0}^{k-1}(-1)^{k-j}
{k \choose j} N^{j}.
\label{NDOT}
\end{equation}
Although one cannot find $Q_N$ or $Q'_N$ explicitly without solving 
the full exclusion problem, we can take $k=1$ in (\ref{NDOT}) to determine the  
mean wall velocity via
\begin{equation}
{\partial \over \partial t} \langle N \rangle = -w_{-} 
\sum_{N=0}^{\infty} Q'_N+w_{+}.
\label{meanV}
\end{equation}
If the mean wall position is time-independent, $\sum_{N=0}^{\infty}
Q'_N = w_{+}/w_{-}$, and the expected occupancy of the site
immediately preceding the wall is
\begin{equation}
\langle \s_{N-1}\rangle =1-\sum_{N=0}^{\infty}
Q'_N=1-\frac{w_{+}}{w_{-}}. \label{BeforeWall}
\end{equation}
We show in section \ref{unstable} that there are some parameter
regimes in which (\ref{BeforeWall}) cannot be satisfied.  For these
parameter values, there exists no time-independent mean wall
position. However, one can still use (\ref{meanV}) to determine the
relevant mean wall dynamics.  The preceding analysis suggests that it
may be more natural to define sites near the wall by their position
relative to the wall than by their absolute position on the lattice.
To avoid working in both frames of reference, in the next section, we
will begin by considering the limit in which the wall hopping rates
are small compared to other rates in the problem ($p$, $k_+$, and
$k_-$).  In this limit, the wall dynamics are slow compared to the
particle dynamics, and we will assume that the wall frame is
stationary.

\section{Mean Field Solution of Density Profile}\label{SIGMA}

In the $w_{\pm}/k_{\pm}, w_{\pm}/p \rightarrow 0$ limit, we expect the
wall to be nearly stationary.  Mean field equations can be derived by
ensemble-averaging the rate equations for the occupation variables
$\sigma_{i}$, and ignoring correlations ($\langle
\sigma_{i}\sigma_{j}\rangle \approx
\langle\sigma_{i}\rangle\langle\sigma_{j}\rangle$). Upon defining the
mean occupation $s_{i} \equiv \langle\s_{i}\rangle$, the mean field
equations for a {\it fixed} ($w_{\pm} = 0$) wall system in NESS are

\begin{eqnarray}
\displaystyle {\dd s_{i} \over \dd t} =
-s_{i}(1-s_{i+1})+s_{i-1}(1-s_{i}) - k_{-}s_{i} \qquad \nonumber
\\[2pt] +k_{+}(1-s_{i}) = 0, \qquad \label{MFTa} \\[13pt]
\displaystyle {\dd s_{1} \over \dd t} =
\a(1-s_{1})-k_{-}s_{1}-s_{1}(1-s_{2}) \qquad \,\,\, \nonumber \\[2pt]
+k_{+}(1-s_{1})= 0, \qquad\qquad  \,\,\label{MFTb} \\[13pt]
\displaystyle {\dd s_{N-1} \over \dd t} = -k_{-}s_{N-1}
+k_{+}(1-s_{N-1}) \qquad \qquad \qquad \, \nonumber  \\[2pt]\:
+s_{N-2}(1-s_{N-1}) = 0. \qquad \quad 
\label{MFTc}
\end{eqnarray}
where the adsorption, desorption and injection rates have been
normalized by $p$ and time has been rescaled by $p^{-1}$ -- hence,
$k_{\pm}$, $\alpha$ and $t$ in (\ref{MFTa}-\ref{MFTc}) are dimensionless.

However, in order to use condition (\ref{BeforeWall}), we need
expressions for particle density at sites defined by their distance
from the wall.  In the fluctuating  frame of the wall, we use the notation
$\tilde{s}_j \equiv s_{N-j}$.  Upon rewriting (\ref{MFTa}-\ref{MFTc})
in the wall frame, we find
\begin{widetext}
\begin{eqnarray}
\displaystyle \frac{\dd \tilde{s}_j}{\dd t} &=&\displaystyle
-(1+w_{+})\tilde{s}_j (1-\tilde{s}_{j-1}) + (1+(1-\tilde{s}_{1})
w_{-}) \tilde{s}_{j+1} (1-\tilde{s}_{j}) - k_{-}\tilde{s}_{j} +
k_{+}(1-\tilde{s}_{j}) \nonumber \\ \: & \: & \displaystyle
\hspace{4.5cm} -w_-(1-\tilde{s}_1)(1-\tilde{s}_{j+1})\tilde{s}_j +
w_{+}(1-\tilde{s}_{j})\tilde{s}_{j-1} = 0, \label{MFTEQNa} \\[13pt]
\displaystyle {\dd \tilde{s}_{N-1} \over \dd t} &=& \displaystyle
\alpha (1-\tilde{s}_{N-1}) -k_- \tilde{s}_{N-1} -
\tilde{s}_{N-1}(1-\tilde{s}_{N-2})  +k_{+}(1-\tilde{s}_{N-1})= 0, \label{MFTEQNb} \\[13pt]
\displaystyle {\dd \tilde{s}_1 \over \dd t} &=& \displaystyle
-k_{-}\tilde{s}_{1}+k_{+}(1-\tilde{s}_{1})+ (1+w_{-})\tilde{s}_{2}
(1-\tilde{s}_{1}) - \tilde{s}_{1} w_{+} = 0. \label{MFTEQNc}
\end{eqnarray}
\end{widetext}
As expected, (\ref{MFTEQNa}) and (\ref{MFTEQNc}) reduce to
(\ref{MFTa}) and (\ref{MFTc}) in the $w_{\pm} =0$  limit.
If the position of the wall were fixed, we could simply use the
iteration given by (\ref{MFTa}), along with boundary conditions
(\ref{MFTb}) and (\ref{MFTc}) to solve for the density profile $s_i$.

Now consider a moving wall problem. Because $\langle N\rangle$ is
undetermined, we need three conditions to solve (\ref{MFTEQNa}). In
addition to the two boundary conditions (\ref{MFTEQNb}) and
(\ref{MFTEQNc}), we require a third condition,
$\tilde{s}_1=1-w_{+}/w_{-}$, to determine $\langle N \rangle$.  This
third boundary condition fixes $\tilde{s}_1$; $\tilde{s}_2$ is set by
(\ref{MFTEQNc}), and we can use (\ref{MFTEQNa}) to iterate forward in
$j$ as many times as required toward the injection site, until
(\ref{MFTEQNb}) is satisfied.  The number of iterations required to
satisfy (\ref{MFTEQNb}) determines the mean position, $\langle
N\rangle$, of the left boundary, and hence the NESS size reached by the
system.  Although (\ref{MFTEQNa}) was derived in the wall frame, the
resulting density profile is nearly identical to a stationary frame
profile derived from (\ref{MFTa}) when $s_{i}$ is not varying rapidly
with site $i$.  See the Appendix for further discussion.

\begin{figure}[h]
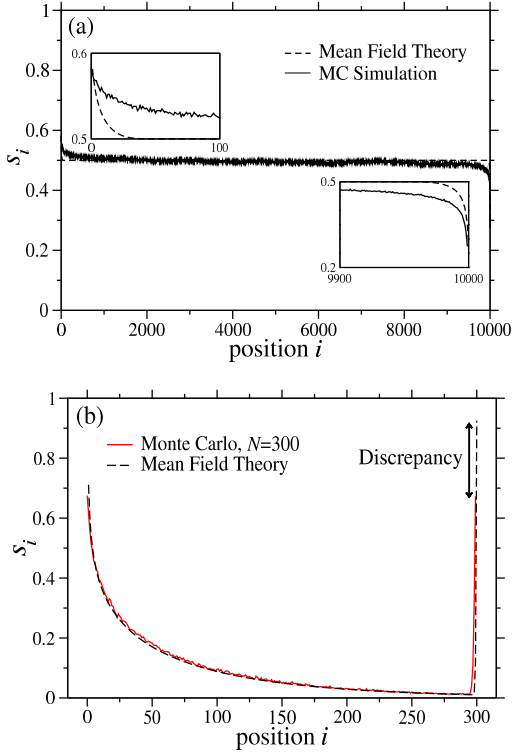

\begin{center}
\includegraphics[height=1.9in]{Fig2a.eps}\\
\vspace{3mm}
\includegraphics[height=1.9in]{Fig2b.eps}
\end{center}
\vspace{-4mm}
\caption{(Color online) A comparison of density profiles derived from Monte-Carlo
simulation and MFT.  (a) The MFT and MC density profiles for conserved
particle TASEP are compared ($\alpha = 0.6$, $\beta = 0.9$, $N =
10,000$).  Despite differences between MFT and MC in the boundary
layers, the particle density at the ends ($i = 1$, $i = 10000$) are
matched through particle conservation.  Insets show the left and right
boundary layers in detail. (b) For a TASEP with Langmuir kinetics
(with $k_- = 0.01$, $N=300$, $p = 1$, and $k_+, w_+, w_- = 0$), the
MFT density profile can be appreciably different from the MC results,
especially near the boundaries.}
\label{DIFFERENCE}
\end{figure}

For standard particle-conserving TASEP models, away from boundaries,
MFT predicts the particle densities to a very high
accuracy \cite{Derrida-92}. A conservation law for the particle density
can be used to fix the end densities to their exact values so that the
MFT also performs well near boundaries \cite{Lakatos-06}.  In Figure
\ref{DIFFERENCE}a, we plot the density profiles from Monte-Carlo
simulations and mean field recursion relations for the simple TASEP
($k_{\pm} = 0$) with a fixed number of sites, $N=10000$.  Differences
in the density profiles are evident in the insets.
%

Because we include particle adsorption and desorption through Langmuir
kinetics, there is no conservation law for the particle density.  In
this case, the boundary densities are not fixed and we see in Fig.
\ref{DIFFERENCE}b that simple mean field calculations of the boundary
density can differ appreciably from the values found from Monte-Carlo
simulations. However, MFT still matches simulation results in the bulk
where $s_i$ varies slowly.  In the following section, we use an
approach that couples explicit enumeration within a finite segment of
sites to the mean field results accurate outside the segment. This
finite segment mean field theory (FSMFT) includes particle
correlations within a segment of sites adjacent to the wall.

\section{Finite Segment Method}\label{sec:FS}

We have shown that mean field theory does a poor job of predicting the
profile $s_{i\approx N-1}$ near the wall when there is a boundary
layer.  To more accurately compute the particle density in this
region, we will solve the Master equation for a finite segment of $m$
sites preceding the wall.  First, we introduce some notation to
explain the mechanics of the finite-segment mean field theory (FSMFT).
For the binary string $(\sigma_{N-m}, ..., \sigma_{N-2},
\sigma_{N-1})$, corresponding to the occupancy of sites in the finite
segment we define the \emph{state} of the segment as the base ten
value of the string.  For example, for $m=2$ sites just left of the
wall, we have four possible combinations for the occupancies (00),
(01), (10), and (11) corresponding to states $i=0,1,2, 3$,
respectively.  If $P_{i}$ is the probability that the finite segment
configuration is in state $i$, the Master equation is
$\partial_{t}P_{i} = M_{ij}P_{j}$ where $M_{ij}$ is the transition
matrix.  In the $m=2$ case,

\begin{widetext}
\begin{equation}
{\bf M} = \left(\begin{array}{cccc}
-(1+w_{-})s^*-2k_{+} & k_{-} & k_{-}+w_{+} & 0 \\[13pt]
k_{+} & -(w_{+}+k_{-}+k_{+}) - s^* & 1 + w_{-}(1-s^*) & k_{-}\\[13pt]
(1+w_{-})s^{*}+k_{+} & w_{+} & -(w_{+}+w_{-}+1+k_{-}+k_{+}) & k_{-}+w_{+} \\[13pt]
0 & k_{+}+s^{*} & k_{+}+w_{-}s^{*} & -2k_{-}-w_{+}\end{array} \right)
\end{equation}
\end{widetext}
where $s^*\equiv \langle \sigma_{N-m-1}\rangle$ is the mean occupancy
in the lattice site just to the left of the segment.  The mean
occupancies in the finite segment can be calculated from ${\bf M}$ in
the following way.  First, the eigenvector, ${\bf P}^{(0)}$,
corresponding to the eigenvalue zero is computed.  The vector ${\bf
P}^{(0)}$, normalized such that $\sum_{i}^{m} P_{i}^{(0)} = 1$
corresponds to the stationary probability distribution, {\it i.e.},
$\partial_t {\bf P}^{(0)} = 0$.  Let $\textbf{v}$ be a $m \times
2^{m}$ matrix where the columns are the ordered state vectors.  The
mean densities are then given by
$(s_{N-m-1},\ldots,s_{N-2},s_{N-1})^{T} = {\bf v}{\bf P}^{(0)}$.

\begin{figure}[h]
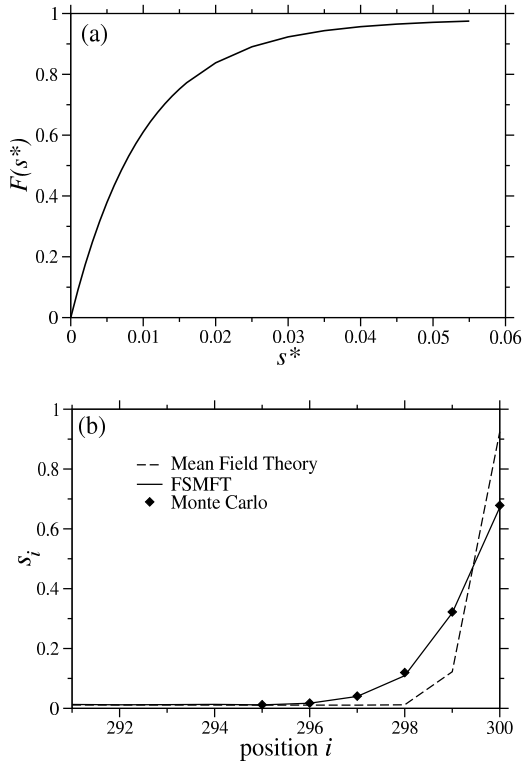

\begin{center}
\includegraphics[height=1.9in]{Fig3a.eps}\\
\vspace{4mm}
\includegraphics[height=1.9in]{Fig3b.eps}
\end{center}
\vspace{-4mm}
\caption{(a) $F(s^*)$ is plotted with parameter values $k_+ = 0$, $k_-
= .01$, $w_+ = 0.005$, $w_- = 0.01$.  (b) The finite segment method
predicts the boundary layer profile significantly better than mean
field theory.  Here, the final ten sites of a fixed-wall profile
($N=300$) are plotted.  The parameters $k_-=0.01$, $k_+ = 0$, and
$\alpha=1$ were used.}
\label{SEGMENT}
\end{figure}

For every value of $s^*$, FSMFT can be used to compute the
mean densities $s_{N-m}, ..., s_{N-2}, s_{N-1}$. In particular, it
establishes a one-to-one correspondence between $s^*$ and $s_{N-1}$:

\begin{equation}
s_{N-1} = F(s^*;w_{\pm},k_{\pm}). \label{FSM-relation}
\end{equation}

\noindent Our calculations indicate that $F(s^*)$ is always a monotonically
increasing function of $s^*$, as shown in Figure \ref{SEGMENT}a.
Comparing the density profiles near a {\it fixed} wall
($w_{+}=w_{-}=0$), Figure \ref{SEGMENT}b shows that using FSMFT (with
$m=5$) significantly improves our prediction of the particle density
near the wall over that obtained using simple ($m=1$) MFT.  To
calculate $\langle N \rangle$ for a fluctuating wall ($w_{-} > w_{+}
>0$), we first solve for the profile of a segment of sites adjacent to
the wall.  From (\ref{BeforeWall}), when the wall attains a
steady-state position, $s_{N-1}=1-w_{+}/w_{-}$.  Using
(\ref{FSM-relation}), we find the value of $s^*$ satisfying
$1-w_{+}/w_{-} = F(s^*)$.  Defining this particular value of $s^*$ as
$s^*_{eq}$, we then use the recursion relation given by
(\ref{MFTEQNa}) to solve the density profile to the left of the finite
segment.  Using the values of $s_{N-m}$ and $s^*_{eq} \equiv
s_{N-m-1}$ from the FSMFT as starting conditions for the recursion
equations, we iterate to the left until the left boundary condition
(\ref{MFTEQNb}) is satisfied.  In summary, the finite segment mean
field theory (FSMFT) is implemented by the following steps:

\vspace{4mm}

$\bullet$ For a given $s^{*}$, solve for the normalized eigenvector
corresponding to the zero eigenvalue of the $2^{m}\times 2^{m}$
transition matrix $M_{ij}(s^{*})$.


\vspace{3mm}

$\bullet$ From the zero eigenvector, express the mean density $s_{N-1}$
 at the site nearest the wall as a function of $s^{*}$, giving relation
 (\ref{FSM-relation}).

\vspace{3mm}

$\bullet$ For a static wall  NESS, set $s_{N-1}=1-w_{+}/w_{-}$ and find
$s_{eq}^{*}$ that yields zero net wall drift by using $1-w_{+}/w_{-} =
F(s_{eq}^*;w_{\pm},k_{\pm})$.

\vspace{3mm}

$\bullet$ Starting with $\tilde{s}_{m+1} = s_{eq}^{*}$ (and
$\tilde{s}_{m} = s_{N-m}$) iterate using the simple mean field
equation (\ref{MFTEQNa}) until equation (\ref{MFTEQNb}) is satisfied.

\vspace{3mm}

$\bullet$ The number of iterations required determines the mean wall
position ($\langle N\rangle \approx$ number of iterations $+m+2$) as a
function of the rate parameters through the starting value
$s_{eq}^{*}$.

\vspace{4mm}

We expect the predicted results from a moving-frame FSMFT to be in
good agreement with those from MC simulations. This is because in
regions where $s$ is slowly varying, the mean field equations
describing the density profile in the wall frame and in the lab frame
yield nearly identical profiles. This can be seen from the continuum
equations, as will be discussed in the Appendix. In these regions,
accurate estimates of the mean wall position can also be obtained
using the continuum approximations to (\ref{MFTEQNa}), provided
$\langle N \rangle$ is large.  When state enumeration of a larger
segment is used, more of the correlations within the density boundary
layer is taken into account and more accurate results are
expected. Provided most of the regions with large gradients in density
is captured by the finite segment, the results will be very accurate.
The incremental accuracy achieved as larger segments are used has been
discussed in a different, but related system \cite{Chou-04}. In the
subsequent analyses, we use a five-site ($m=5$) FSMFT -- generating a
$2^{5}\times 2^{5}$ eigenvalue problem in the process -- and
self-consistently solve for the densities away from the boundary
layer.  This choice of segment size is sufficient to yield accurate
results for all parameters explored.

\section{Results and Discussion} \label{RESULTS}

\subsection{Time-Independent Mean Wall Positions} \label{NO-KON}
We first consider regimes in which the wall acquires a
static mean position in NESS. Using Monte Carlo simulations
and FSMFT, we study the dependence of the mean wall position on the
injection rate $\alpha$, particle adsorption and desorption rates
$k_{\pm}$, and the wall hopping rates $w_{\pm}$.
We can use analytic solutions of the bulk continuum equations in order
to understand parameter dependencies of our model.  Although mean
field theory poorly describes our system in the boundary layers where
the profile varies rapidly, away from boundary layers, simple MFT is
accurate.  In these regions, to guide our our analysis, we will use
the continuum limit of the mean field equations.  We define
$\varepsilon\equiv 1/N_0$ where $N_0$ is a characteristic number of
lattice sites (to be derived below) and $x \equiv (i-1)/N_0 $ as a
relative position along the lattice.  As shown in the Appendix, the
NESS density profile obeys

\begin{equation}
\varepsilon(2s-1)s'(x)+k_+-(k_+ + k_-)s + O(\varepsilon^2) = 0
\label{eq:bulk1}
\end{equation}
in both the lab and wall frames of reference.  Upon integrating, we
obtain the implicit equation
\begin{equation}
\begin{array}{l}
\displaystyle \frac{(k_+ -k_-)\ln\vert k_{+}-(k_{+}+k_{-})s\vert 
-2k_{+} + 2(k_{+} + k_{-})s}{(k_{+}+k_{-})^2} \\[13pt]
\hspace{5cm} \displaystyle ={x\over \varepsilon}+C,
\end{array}
\label{bulk}
\end{equation}
where $C$ is a constant of integration.  In the continuum description,
the entrance site is at position $x = 0$, the wall's position is $L$,
and the mean wall position is $\langle L \rangle \equiv \varepsilon
\langle N \rangle$.  We can use (\ref{bulk}) to understand the
behavior of the mean wall position $\langle N \rangle$.  First, note
that the left hand side of (\ref{bulk}) scales as
$(k_++k_-)^{-1}$. Since $\varepsilon^{-1} \equiv N_0$ scales as
$(k_-+k_+)^{-1}$, we define $N_0 \equiv (k_+ + k_-)^{-1}$.  For a
continuum description to be useful, $N_0$ must be large, so $k_{\pm}$
must be small.

\begin{figure}[t]
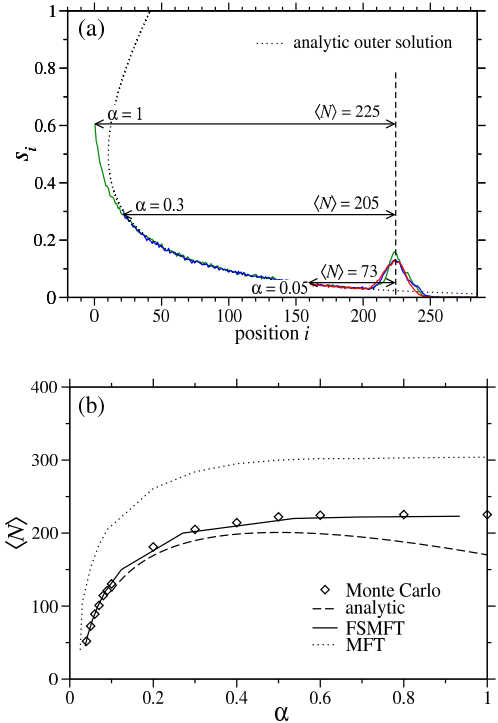

\begin{center}
\includegraphics[height=1.8in]{Fig4a.eps}\\
\vspace{4mm}
\includegraphics[height=1.8in]{Fig4b.eps} 
\end{center}
\vspace{-4mm}
\caption{(Color online) Simulations were performed with $w_+=0.001$, $w_-=0.01$,
$k_-=0.01$, and different values of $\alpha$.  In (a), profiles for
three simulations are plotted with their mean wall positions aligned.
When $\alpha \lesssim 0.5$, $s(0)\approx \alpha$. When $\alpha \gtrsim 0.5$, $s_B$
becomes multivalued and there is a boundary layer on the left.  Within
the boundary layer, a small change in the position results in a large
change in the particle density.  Thus, in (b), where $\langle N
\rangle$ is plotted as a function of $\alpha$, large changes in
$\alpha$ result in small changes in $\langle N \rangle$ when
$\alpha\gtrsim 0.5$. Also shown is the prediction from (\ref{meanN}) with
$s_{eq}^{*} = 0.026$, determined by FSMFT.} \label{fig:Alpha}
\end{figure}
 
Equation (\ref{bulk}) gives an implicit formula for the bulk density,
which we denote by $s_B(x)$, in terms of the adsorption and desorption
rates, $k_{\pm}$ and the integration constant $C$.  The injection rate
$\alpha$ determines $C$, and along with the wall hopping rates
$w_{\pm}$, determines the mean wall position. As shown in the Appendix,
the solution near the left boundary varies slowly when $\a \lesssim
0.5$. Furthermore, if $k_- \ll \alpha$, we can approximate $s_1
\approx s_2$ in (\ref{MFTb}) to conclude that $s(0) \approx \alpha$.
This simplified condition can be used to determine $C$ in
(\ref{bulk}).

When $\alpha \gtrsim 0.5$, a boundary layer arises on the left (this can be
seen in Fig. \ref{fig:Alpha}a).  In this regime, $s(0)$ can no longer
be approximated as $\alpha$, and $s_B(x)$ becomes invalid near the
injection site.  While $s_B(x)$ is still a good approximation to the
density profile outside the boundary layer (where $s\lesssim  0.5$), there is
no straightforward, analytic way to calculate $C$ when $\alpha\gtrsim 0.5$.
In Fig. \ref{fig:Alpha}a, when $\alpha = 1$, $C$ is used as a single
fitting parameter and is determined empirically such that in the bulk,
$s_B(x)$ approximates the density profile obtained using MC
simulations. The mean wall position, $\langle L \rangle$, is found
through the relation $s_B(\langle L \rangle-(m+1)) = s^*_{eq}$ where
$s^*_{eq}$ is found using an $m-$site FSMFT and is the value of $s^*$
that puts no net drift on the wall.


Figure \ref{fig:Alpha}a shows results from MC simulations, shifted so
that the mean wall positions are aligned at $\langle N \rangle = 225$,
which is the mean wall position when $\alpha = 1$.  While the density
profile has a sharp boundary layer at the wall in the wall frame, in
the lab frame, the boundary layer is smeared out due to wall
fluctuations.  This results in the broad peaks centered on the mean
wall position seen in Figure \ref{fig:Alpha}a.  The outer solution
$s_B(i/N_0)$ with $N_0 = (k_+ + k_-)^{-1}$ is shown by the dotted
curve.  The close agreement between the MC data and $s_B(x)$ suggests
that dropping the $O(\ve^2)$ term in (\ref{eq:bulk1}) to obtain
$s_B(x)$ produces an excellent approximation to the mean particle
density, provided $\a \lesssim 0.5$.  Note that $\alpha$, through $C$,
simply shifts $s_B(x)$ to the left or right; thus, when we vary only
$\alpha$ and plot the resulting density with the mean wall positions
aligned (as they are in \ref{fig:Alpha}a), the profiles collapse onto
the same curve.

We can also use (\ref{bulk}) to predict the mean wall position as a
function of the injection rate $\a$ when $\a$ is not too large.  For
simplicity, consider $k_+ = 0$ -- the analysis for $k_+ \neq 0$ is
analogous.  Using the simplified condition $s(0) = \alpha$ in
(\ref{bulk}), we have
\begin{equation}
C = {2\alpha -\ln (\alpha k_{-}) \over k_{-}}.
\label{eq:C}
\end{equation} 
Now, using the relation $s^*_{eq}=s_B(\langle
 L\rangle-\varepsilon(m+1))$ and (\ref{eq:C}), (\ref{bulk}) becomes
\begin{equation} 
\langle N\rangle = \frac{1}{k_-} \ln \left( \frac{\alpha}{e^{2\alpha}}
\frac{e^{2 s^*_{eq}}}{s^*_{eq}}\right)+m+1.
\label{meanN}
\end{equation}

The dependence of $\langle N \rangle$ on $\a$ is shown in Figure
\ref{fig:Alpha}b, predicted using four different methods.  Simple MFT
($m=1$, dotted curve) performs poorly relative to MC simulations (open
diamonds). The results from FSMFT with $m=5$ (solid curve) agree very
well with the MC data for all values of $\alpha$.  The solution of
(\ref{meanN}) (dashed curve) performs reasonably well provided $\a$ is
not too large. When $\a \gtrsim 0.5$, $s(0) = \alpha$ is a poor approximation
to (\ref{MFTb}) and the resulting prediction of $\langle N \rangle$
suffers. In fact, the slope $s_B'$ diverges when $s_B=0.5$, which can
be seen from (\ref{eq:bulk1}).  When $\alpha \gtrsim 0.5$, there is a
boundary layer on the left with width $O(\sqrt{\ve})$
(cf. Appendix). As a result, increases in $\alpha$ above $0.5$ will
increase the height of the boundary layer, but will not significantly
change the mean wall position, and $\langle N \rangle$ becomes
insensitive to changes in $\alpha$ \footnote[2]{In \cite{Evans-03,Frey-04}, 
an asymmetric exclusion process in a fixed
domain with open boundaries and Langmuir kinetics was studied. 
Regimes arise in which the position of a shock in the particle density
becomes insensitive to the ejection rate $\beta$ on the right, once 
$\beta >0.5$.  Because these observations were made
under the assumption $k_+>k_-$, and we consider $k_->k_+$, by particle
hole symmetry, the ejection rate in these works corresponds to
the injection rate, $\alpha$ in our problem.  While the authors of
\cite{Evans-03, Frey-04} find a shock position insensitive to boundary
conditions, we find an insensitive mean wall position.}.

\begin{figure*}[htb]
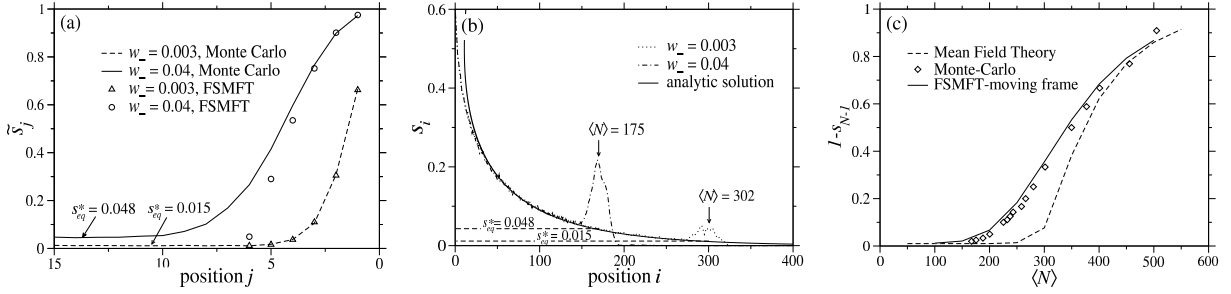

\begin{center}
\includegraphics[height=1.5in]{Fig5a.eps}
\hspace{1mm}
\includegraphics[height=1.5in]{Fig5b.eps}
\hspace{1mm}
\includegraphics[height=1.5in]{Fig5c.eps}
\end{center}
\vspace{-4mm}
\caption{Boundary effects near the wall determine wall position.  In
(a), $s^*_{eq}$ is determined empirically from MC simulations captured
in the wall frame and numerically using the finite segment method.  In
(b), MC simulations are plotted in the lab frame.  In (c), the
occupancy of the last site in the wall frame $s_{N-1} = 1-w_{+}/w_{-}$
is plotted as a function of the mean wall position $\langle N
\rangle$.  The parameters $\alpha =1$, $k_-=0.01$, $k_+ = 0$, and $w_+
= 0.001$ were used.}
\label{fig:wp}
\end{figure*}

We now discuss how changes in the wall hopping rates can affect the
wall position. In Fig. \ref{fig:wp}a, for a fixed value of $w_{+}$,
one sees that an increase in $w_{-}$ increases the value of
$s^*_{eq}$. Our FSMFT predicts that given values of $w_+$ and $w_-$,
$s^*_{eq}$ must satisfy $1 - w_{+}/w_{-} = F(s^*_{eq};w_{\pm})$. 
For small values of $w_{\pm}$, $F(s_{eq}^{+},w_{\pm})\approx 
F(s_{eq}^{*},0) \approx 1-w_{+}/w_{-}$, suggesting that
$s^*_{eq}$ depends primarily on the ratio $w_+/w_-$, with only a weak
dependence on the individual wall hopping rates.  Since
$F$ is a monotonically increasing function, 
$s^*_{eq}$ increases with $w_{-}/w_{+}$.

A change in $s_{eq}^{*}$ induces a change in the mean wall position,
shown in Fig. \ref{fig:wp}b. Again, this is consistent with our theory
since $\langle L \rangle$ must satisfy $s_B(\langle L\rangle
-\ve(m+1)) = s_{eq}^{*}$. In the special case $\alpha \lesssim 0.5$,
one can use (\ref{meanN}) to predict $\langle N\rangle$ directly,
given $s^*_{eq}$. When $\alpha \gtrsim 0.5$, one either has to solve the
full set of discrete MFT equations (\ref{MFTEQNa}) and (\ref{MFTEQNb})
-- or the equivalent continuum equations (\ref{CONTINUUM}) and
(\ref{BCLEFT}) -- coupled to a finite segment, to obtain $\langle
N\rangle$. Our results from solving the discrete equations are shown
in Fig. \ref{fig:wp}(c). Using simple MFT ($m=1$) without a larger
finite segment generally results in poor predictions for $\langle
N\rangle$.



A more complete understanding of the wall dynamics can be garnered by
analyzing the wall fluctuations.  For simplicity, we consider the
continuum description in which the wall's motion can be approximately
described by a diffusion constant $D = \varepsilon^2 w_+$ and a
position-dependent drift, $V(L)$.  If one assumes that the wall
fluctuates within a harmonic ``potential,'' this drift takes the form
$V(L) = -a(L-\langle L \rangle)$, where $a \equiv -(\dd V/\dd
L)|_{\langle L \rangle}$.  This approximation effectively closes
(\ref{EFFECTIVEQ}) by expressing the effects of conditional
probability, $Q_{N}'$, in terms of a drift. For $L \approx \langle
L\rangle$, the probability density of the wall's position, $Q(L)$ (the
continuum analog of $Q_N$), can be approximately found from the
solution of
\begin{equation}
\frac{\partial Q(L,t)}{\partial t} = a{\partial \over \partial L}\left[
(L-\langle L\rangle)Q(L)\right]+ D\frac{\partial^2 Q}{\partial
L^2}.
\label{eq:Diffusion}
\end{equation}
Upon imposing the normalization $\int_{-\infty}^{\infty}Q(L)dL = 1$, we find the 
steady-state solution to (\ref{eq:Diffusion}):
\begin{equation}
Q(L)=\sqrt{\frac{a}{2D\pi}}e^{-\frac{a(L-\langle L \rangle)^2}{2D}},
\label{eq:Gauss}
\end{equation}
where $a$ is given by
\begin{equation}
a \equiv -\frac{d V}{d L}\bigg|_{\langle L \rangle} = -\frac{\partial
V}{\partial F}\frac{\partial F}{\partial s^*}s_B'(\langle L \rangle -
\varepsilon(m+1)).
\label{A}
\end{equation}

The drift velocity $V(L)$ can be inferred from $\ve(w_{+}-w_{-}(1-s_{N-1}))$
and (\ref{FSM-relation}), which relates the mean occupancies at positions $L$ and
$L-\ve(m+1)$,

\begin{equation}
V(L) = \ve\left[w_{+} - w_{-} + w_{-} F(s^*(L-\varepsilon(m+1)))\right].
\label{wall-vel}
\end{equation}

\noindent By defining the drift $V(L)$ using the steady-state relation
$F$, we have implicitly made an adiabatic approximation where the
particles have reached a NESS for any wall position $L$.  

\begin{figure}[h]
\begin{center}
\includegraphics[height=1.9in]{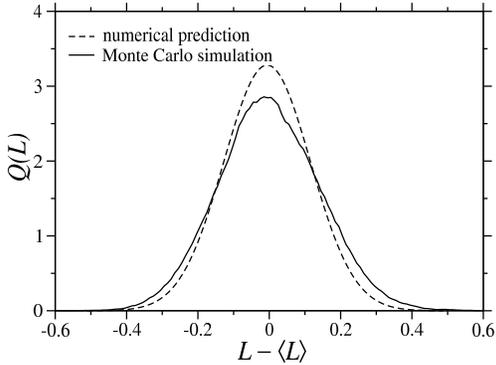}
\end{center}
\vspace{-4mm}
\caption{Probability density of the wall position, $Q(L)$, is plotted
as a function of the deviation from the mean wall position $\langle
L\rangle$.  The parameters $\alpha = 1$, $w_{+} = 0.005$, $w_{-} =
0.01$, $k_- = 0.01$, and $k_+ = 0$ yield $\langle L \rangle \approx
350$.  The distribution predicted from (\ref{eq:Gauss}) is a close
approximation to that derived from MC simulations.}
\label{fig:WallDist}
\end{figure}

We can also estimate the variance of the wall position by using
$\Sigma^{2} \approx D/a$, and (\ref{A}) for $a$. Upon differentiating
$V(L) = \varepsilon(w_+-w_-[1-F(s_B'(L- \varepsilon(m+1)])$, we find
$(\partial V/\partial F) = \varepsilon w_-$.  We can estimate
$(\partial F/\partial s^*)\vert_{s^{*}_{eq}}$ using the finite segment
method, and we know $s_B'(\langle L \rangle - \varepsilon(m+1))$
exactly from (\ref{eq:bulk1}).  Assuming that $\langle L^2 \rangle
\approx \int_{-\infty}^{\infty} dL L^2 Q(L) $, we expect the variance
of the wall position to be approximately
\begin{equation}
\Sigma^2=\langle L^2 \rangle - \langle L \rangle^2 =
\frac{D}{\varepsilon w_- F'(s^*_{eq}) s_B'(\langle L
\rangle - \varepsilon(m+1))}
\label{eq:Variance}
\end{equation}
In Figure \ref{fig:WallDist}, we plot an example distribution $Q(L)$
found using both Monte Carlo simulations and from (\ref{eq:Gauss}).
We have aligned the distributions such that their maxima coincide.
Using (\ref{eq:Variance}), the standard deviation $\Sigma \approx
0.1215$, in good agreement with the standard deviation found from MC
simulations $\Sigma\approx 0.143$. Since $\Sigma/\langle L\rangle
\sim 0.14/3.5 \ll 1$, the wall is fairly stable and not likely to fall
off the injection end of the lattice except on exponentially long time
scales.

\subsection{Time-dependent mean wall positions}
\label{unstable}
In the previous section, we explored the dependence of the
statistically stationary mean wall position on the model
parameters. However, a stable mean wall position may not always
exist. In this section, we use a FSMFT to determine the stability of
the wall and the conditions under which a permanent net wall drift
might arise.

\begin{figure}[h]
\begin{center}
\includegraphics[height=2.2in]{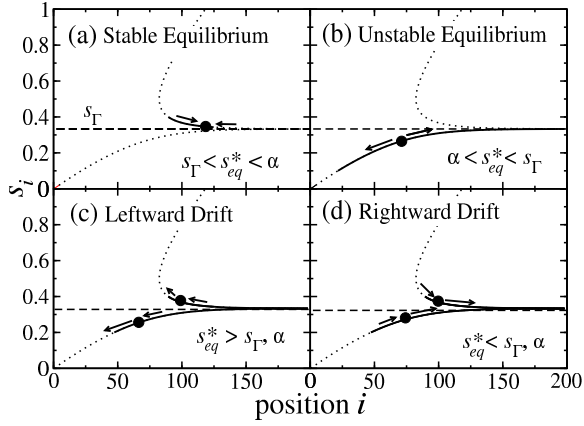}
\end{center}
\vspace{-4mm}
\caption{Four possibilities for the wall motion. Arrows indicate the
mean wall motion if it is at position $i$.  (a) the wall could have no
net drift and a stable fixed mean position, with small perturbations
to the wall position decaying over time, (b) the wall could have no
net drift, but an unstable fixed mean position, with small
perturbations to its position causing it to drift indefinitely to the
left or right, (c) the wall could drift indefinitely to the left and
(d) the wall could drift indefinitely to the right. The outer solution
$s_{B}(x)$ is shown by the dotted curves.}
\label{fig:dynamics}
\end{figure}

The motion of the wall can be understood completely in terms of the
outer solution $s_B(x)$ -- given by inverting (\ref{bulk}) -- and the
particle density inside finite segment. First, we consider some
important properties of $s_B(x)$. Equation (\ref{bulk}) admits two
branches to the bulk solution, $s_B(x)$, because the argument of
$\ln[|\quad|]$ can either be positive or negative.  The argument
approaches zero as $s$ approaches $s_{\Gamma} \equiv
k_{+}/(k_{+}+k_{-})$, the density arising from Langmuir kinetics
alone.  When we invert $x(s_B)$ to find $s_B(x)$, we see that for
increasing $x$, one branch of the density profile $s_B(x)$ approaches
$s_{\Gamma}$ asymptotically from below, and a second branch approaches
$s_{\Gamma}$ asymptotically from above. A representative $s_B(x)$ is
plotted in Figure \ref{fig:dynamics}.  Notice that $s'(x)>0$ in the
lower branch, and $s'(x)<0$ in the upper branch
\footnote[3]{When $s_B(x)$ passes though $s_B(x) = 0.5$, the sign of
$s'_B(x)$ changes.  The profile $s(x)$ departs from $s_B(x)$ near
$s_B(x)=0.5$ because one of the assumptions used to derive $s_B(x)$ --
that $s'_B(x) = O(1)$ -- becomes invalid and the sign of $s'(x)$
fails to change upon passing through $s(x)=0.5$.  See the Appendix for
further discussion of when the continuum equations may cease to be
valid.}.
If $\alpha > s_{\Gamma}$, the steady state density profile will lie on the
upper branch and the bulk density will have values satisfying
$s_{\Gamma}<s_B(x)<\alpha$.  If the injection rate $\alpha < s_{\Gamma}$, the steady
state density profile will lie on the lower branch and the bulk
density will attain values $\alpha<s_B(x)<s_{\Gamma}$.

We are now ready to derive conditions for the existence of a fixed
mean wall position and stability criteria.  In the adiabatic
approximation (\ref{wall-vel}), $m$ is the number of sites in the
finite segment and $\varepsilon = 1/N_0$. This equation expresses the
velocity of the wall at position $\ve N = L$ in terms of a particle
density at position $(N-m-1)\varepsilon$.  Since $s^*_{eq}$ is the
value of $s^*$ that puts no net drift on the wall, {\it i.e.}, $w_+ -
w_-(1-F(s^*_{eq})) = 0$, we can expand $V(L)$ from (\ref{wall-vel}) in
a Taylor Series about $s^*_{eq}$ to find

\begin{equation}
V(L) \approx \ve w_- F'(s^*_{eq})(s^*(L-\varepsilon(m+1)) - s^*_{eq}).
\label{wall-vel2}
\end{equation}
Because $F$ is a monotonically increasing function and $w_{-}> 0$, the
wall drifts to the right if $s^*(L-\varepsilon(m+1)) > s^*_{eq}$, to
the left if $s^*(L-\varepsilon(m+1)) < s^*_{eq}$ and has a fixed mean
position if $s^*(L-\varepsilon(m+1)) = s^*_{eq}$. If we now assume
that $m$ is sufficiently large so that the point $L -
\varepsilon(m+1)$ lies outside of the boundary layer, then
$s^*(L-\varepsilon(m+1))$ can be well approximated by the outer
solution given by $s_B(x)$, i.e. $s_B(L-\varepsilon(m+1)) \approx
s^*(L-\varepsilon(m+1))$.  Furthermore, we know that $s_B(x)$
satisfies $\alpha \leq s_B \leq s_{\Gamma}$ on the lower branch and $s_{\Gamma} \leq
s_B \leq \alpha$ on the upper one. Therefore, we conclude that if
$s^*_{eq} \notin [\alpha, s_{\Gamma}]$, the wall can never have a fixed mean
position.  In particular, for all $t$
\begin{equation}
\begin{array}{rl}
V(L(t)) &  > 0 \quad \mbox{if} \quad  s^*_{eq} < \alpha, s_{\Gamma}, \nonumber  \\
\: &           < 0 \quad \mbox{if} \quad  s^*_{eq} > \alpha, s_{\Gamma},
\end{array}
\end{equation}
corresponding to an indefinite rightward and leftward drift,
respectively (cf. Fig. \ref{fig:dynamics}c and d).
If a fixed mean position does exist, we can understand
its stability by considering the sign of $dV/dL$. If this quantity is
negative (positive), the position is stable (unstable). These
possibilities are summarized in Figure \ref{fig:dynamics}.  By
differentiating (\ref{wall-vel2}), we have

\begin{equation}
{\dd V\over \dd L} \approx \ve w_- F'(s^*_{eq}) s_B'(L-\varepsilon(m+1)).
\end{equation}
Hence, if a mean wall position exists at $L$, a necessary and
sufficient condition for its stability is
\begin{equation}
s_B'(L-\varepsilon(m+1)) < 0.
\label{stab}
\end{equation}
In particular, when there is no adsorption ($k_+ = 0$), the bulk
solution $s_B(x)$ decreases monotonically from the injection site and
any mean wall position $\langle L\rangle$ induced by the kinetics will
be deterministically stable.

\begin{figure}
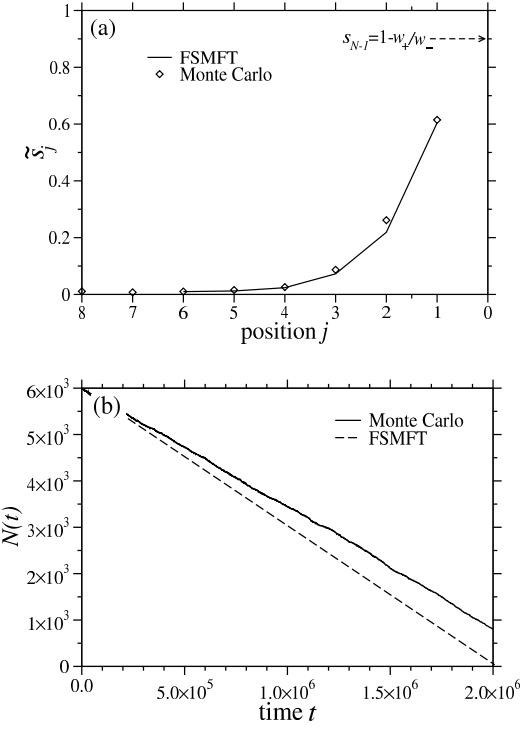

\begin{center}
\includegraphics[height=1.8in]{Fig8a.eps}\\
\vspace{4mm}
\includegraphics[height=1.8in]{Fig8b.eps}
\end{center}
\vspace{-4mm}
\caption{In (a), we plot the density profile near the wall from
FSMFT and from MC simulations.  We find that $s_{N-1}$ is less than
$0.9=1-w_{+}/w_{-}$, the value required for the wall to have zero
drift.  In (b), the position of the wall $N(t)$ found from MC
simulations is plotted with the expected $N(t)$ calculated using
FSMFT.  Parameter values were $\alpha = 0.01$, $w_+=0.001$, $w_-=0.01$,
$k_+=0.0001$, $k_-=0.01$.}
\label{WALLOFF}
\end{figure}  

\begin{figure*}
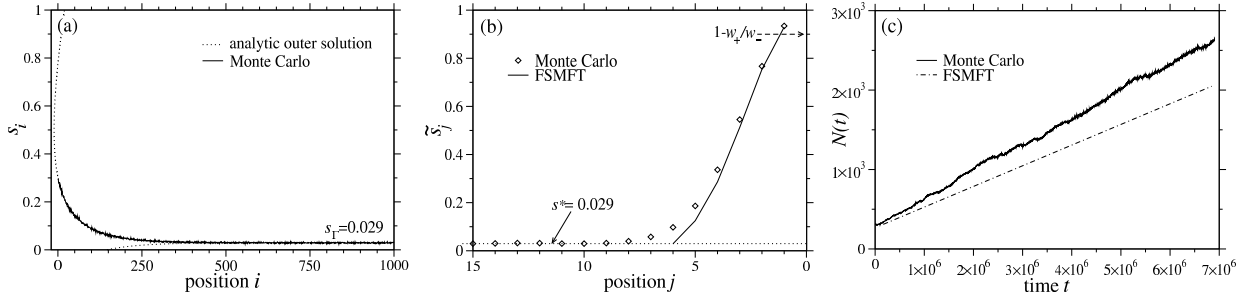

\begin{center}
\includegraphics[height=1.52in]{Fig9a.eps}\hspace{1mm}
\includegraphics[height=1.52in]{Fig9b.eps}\hspace{1mm}
\includegraphics[height=1.52in]{Fig9c.eps}
\end{center}
\vspace{-2mm}
\caption{When $s_{\Gamma}>s^*_{eq}$ and $\alpha>s^*_{eq}$, the wall
escapes.  In (a), far from the injection site, the particle density
approaches $s_{\Gamma}= 0.029$ as predicted by analytic theory.  In
(b), we use FSMFT and MC to find $s_{N-1}>0.9=1-w_{+}/w_{-}$, the
value for which the wall's drift would be zero.  In (c), we show
$N(t)$ to compare the escape velocity calculated from finite segment
analysis to the escape velocity found in simulations.  Although the
value of $s_{N-1}$ found by FSMFT differs from that found in MC
simulations by only $0.4\%$, the calulated velocities differ by
$17\%$.  Parameter values were $\alpha = 0.3$, $w_+ = 0.001$, $w_- =
0.01$, $k_+=0.0003$, and $k_-=0.01$.}
\label{ESCAPE}
\end{figure*}

Figures \ref{WALLOFF} and \ref{ESCAPE} compare the result
(\ref{wall-vel}) with simulation data.  Figure \ref{WALLOFF} shows the
results of a MC simulation in which the wall particle has a mean
leftward drift.  In Fig. \ref{WALLOFF}a, the density profile in the
wall frame found from MC simulation and that predicted using FSMFT are
shown.  Far from the injection site $s_{B}$ asymptotes to
$s_{\Gamma}$. Thus, when the wall starts at a position $L_{0} \gg 1$,
we assume $s^*(L-\varepsilon(m+1))=s_{\Gamma}$ and the wall's velocity
$V$ is independent of its position $L$.
Since $s_{N-1}<1-\frac{w_+}{w_-}$, we expect, from (\ref{wall-vel}),
that the net drift on the wall will be negative.  In
Fig. \ref{WALLOFF}b, we compare $N(t)=L(t)/\ve$ found from MC
simulations with that calculated assuming $L(t) = L_0+V t$ where
$V$ is calculated using (\ref{wall-vel}) in the large $L$ limit.  Similarly,
Fig. \ref{ESCAPE}a shows a density profile from MC simulation in the
case where the wall acquires a mean rightward drift.  In
Fig. \ref{ESCAPE}b both MC simulations and FSMFT show that in the wall
frame, the occupancy of the site adjacent to the wall is greater than
$1-\frac{w_+}{w_-}$, and we expect a mean rightward drift.  In
Fig. \ref{ESCAPE}c, we see that this is the case, and the predicted
time course $N(t)=L(t)/\ve$ is compared with $N(t)$ found using MC
simulations.

In contrast to the case of a static mean wall position, when the wall
has a position-independent velocity, $V$, the diffusion constant $D$
of the wall is given by $D = \varepsilon^2(w_+-w_-+w_-F(s_{\Gamma}))/2$.  The
probability density $Q(L,t)$ describing wall position then follows
\begin{equation}
\frac{\partial Q}{\partial t}=D\frac{\partial^2 Q}{\partial L^2}-V \frac{\partial Q}{\partial L},
\end{equation}
the solution of which is 
\begin{equation}
Q(L,t)=\frac{1}{2\sqrt{\pi D t}}e^{-\frac{(L-Vt-L_0)^2}{4Dt}}. 
\end{equation}

We now discuss our results in the context of the phase transitions
\cite{Evans-03,Frey-04,SCHUTZ2} of the interior density. When Langmuir
kinetics is coupled to a fixed domain TASEP with open boundaries,
qualitative properties of $s(x)$ can change abruptly when
adsorption/desorption and injection/ejection rates vary. For example,
an interior boundary layer separating regions of low and high density
can suddenly disappear, replaced with a single region of high density
as the injection rate $\a$ is increased. 

Our moving wall TASEP system coupled with Langmuir kinetics does
\emph{not} support the phase structure seen in
\cite{Evans-03,Frey-04,SCHUTZ2}.  Because we limit ourselves to
$k_+<k_-$, we can see from (\ref{eq:bulk1}) that when $s_B(x)>0.5$,
(corresponding with a high density region), $s_B'(x)>0$.  From
(\ref{stab}), the wall cannot have a stable equilibrium position
within the high density region, and we do not find time-independent
density profiles with low to high density interior shocks (a low-high
shock), as is observed in \cite{Evans-03,Frey-04}.  In fact,
references \cite{Frey-03,SCHUTZ2} show that high-low shocks are never
stationary in an exclusion process with Langmuir kinetics.  Therefore,
interior shocks are never stable in our model system.  In our problem,
the presence of a wall that responds to particle dynamics relaxes any
shocks in density that may otherwise occur in the interior, forcing
them to the left or right boundaries.

\section{Summary and Conclusions}

Our model of an asymmetric exclusion process with Langmuir kinetics
and a movable right boundary, and the corresponding results provide a
guide to understanding biophysical processes in which many processing
molecular motors push against a load. The detachment and attachment
rate of the motors, as well as the injection rate at the entry site,
determine the load the motors can support. If a static load particle
position is reached, we see that the mean wall distance from the
injection site saturates upon increasing injection rate $\alpha$ past
about $0.5$. The analyses can be used to predict whether biological
processes such as ribosome movement and filopodia/filament extension
continues or reaches a static configuration.

Within our model, we found four parameter regimes.  In the first
regime, $(s_{\Gamma} < s_{eq}^{*} < \alpha)$, the wall attains a
stable equilibrium position for the wall.  In the second regime,
$(\alpha < s_{eq}^{*} < s_{\Gamma})$, there is an equilibrium, but
unstable mean wall position.  In the third and fourth regimes,
$(s_{eq}^{*}\notin [\alpha, s_{\Gamma}])$, the wall will always feel a
net drift to the to the right and left, respectively. In the latter
case, the wall will fall off the lattice in a time scaling linearly
with the starting position.  When there is a stable equilibrium wall
position, we can find the mean wall position $\langle N \rangle$ as a
function of the particle injection rate $\alpha$, the adsorption and
desorption rates $k_{\pm}$, and the intrinsic hopping rates of the
wall $w_{\pm}$.  Determination of $\langle N \rangle$ requires
accurate evaluation of the particle density near the wall.  Using a
hybrid finite segment/mean field approach in the reference frame of
the fluctuating wall, we accurately determine the particle density
near the wall, and use this to determine the wall's steady-state
position.

When there is no steady state wall position, the finite segment mean
field approach allows us to estimate the steady state velocity of the
wall far from the injection site.  In our analysis, we assumed that
the particle density has reached steady state, thus ignoring the
initial particle density profile and wall position.  Even in regimes
where we expect an equilibrium wall position at steady state, if the
wall is initially near the injection site, and the particle density is
initially very low, we would expect the wall to fall off the lattice
before reaching its equilibrium position.  The times to falling off
the lattice may be treated with extensions of large deviation theory
as suggested by Fig. \ref{fig:WallDist} \cite{LDT}.  A number of
interesting extensions of the free boundary problem arise.  For
example, we expect for certain parameter regimes that slow bottleneck
sites \cite{Chou-04} can attract the fluctuating wall.  These features
and other novel applications to biophysical systems deserve
investigation.

\vspace{3mm} 

This material is based upon work supported under a National Science
Foundation Graduate Research Fellowship.  The authors also acknowledge
support from the NSF through grant DMS-0349195, and the NIH through
grant K25 AI41935.
\section{Appendix}\label{Ap:Continuum}

We can take the continuum limit of (\ref{MFTa}) by defining $x=(i-1) \ve$
where $\ve$ is the lattice spacing.  We find,
\begin{equation}
\displaystyle {\partial s(x,t)\over \partial t} = \ve s'(2s-1)+{\ve^{2}\over
2}s'' -k_{-}s + k_{+}(1-s) = 0. 
\label{CONTINUUM}
\end{equation}

\begin{widetext}
The left hand boundary condition, (\ref{MFTb}) becomes
\begin{equation}
\begin{array}{l}
\displaystyle {\partial s(0,t)\over \partial t} = \a(1-s(0)) - k_{-}s(0)
+k_{+}(1-s(0)) - s(0)(1-s(\ve)) = 0.
\end{array}
\label{BCLEFT}
\end{equation}
When significant changes in the solution near $x=0$ vary over a length
scale that is $> O(\varepsilon)$, this equation is well approximated
by
\begin{equation}
\begin{array}{l}
\displaystyle {\partial s(0,t)\over \partial t} = \a(1-s(0)) - k_{-}s(0)
+k_{+}(1-s(0))  - s(0)(1-s(0)-\ve s'(0)) = 0.
\end{array}
\end{equation}
The right hand boundary condition analogous to (\ref{MFTEQNc}), becomes
\begin{equation}
\begin{array}{l}
\displaystyle {\partial s(L,t) \over \partial t} = -k_{-}s(L)
+k_{+}(1-s(L)) +s(L-\ve)(1-s(L)) 
 +w_{-}s(L-\ve)(1-s(L))-w_{+}s(L) = 0.
\end{array}
\label{BCRIGHT}
\end{equation}
Again, when significant changes in the solution near $x=L$ vary over a
length scale that is $>O(\ve)$, this equation is well approximated by
\begin{equation}
\displaystyle {\partial s(L,t) \over \partial t} = -k_{-}s(L)
+k_{+}(1-s(L))+(s(L)-\ve s'(L))(1-s(L)) 
+ w_{-}(s(L)-\ve s'(L)) )(1-s(L))-w_{+}s(L) = 0,
\label{BCrighteps}
\end{equation}
at the free boundary $L(t)$, where $s(L)$ defines the particle density 
at the position just to the left of the wall.
In the wall frame, the continuum limit of equation (\ref{MFTEQNa}) is
\begin{equation}
\begin{array}{l}
\displaystyle {\partial s(x,t)\over \partial t} = \ve [w_+-(1+w_-(1-s_{N-1}))]s'(1-2s)+{\ve^{2}\over
2}[1+w_++w_-(1-s_{N-1})]s'' -k_{-}s + k_{+}(1-s). 
\end{array}
\label{CONTINUUM_WALL}
\end{equation}
\end{widetext}

In the wall frame, an interior particle shifts to the right when it either hops to the
right, which it does with a (normalized) rate of unity, or when the
wall hops to the left, which it does with rate $w_{-}(1-s_{N-1}) = w_{+}$ in steady
state.  
Similarly, a particle shifts to the left when it hops to the left or
when the wall hops to the right, which it does with rate $w_+$.  In
the bulk, where $s'(x)=O(1)$, the diffusive term is small and can be
neglected.  The only term we retain that depends on hopping rates is
$(w_+ - 1 - w_+)(1-2s)s'$, which is equal to the value of the
corresponding term in the lab frame, $-(1-2s)s'$.  The bulk density is
described in both frames by
\begin{equation}
\ve s'(1-2s)+k_-s-k_+(1-s)=0.
\end{equation}
In our problem, the second order term in (\ref{CONTINUUM}) becomes
important in the right hand boundary layer, and in the left hand
boundary layer when there is one.  On the left, when $s \approx 0.5$,
one cannot assume that $s$ varies slowly.  Making the change of
variables $x = \xi X$ ($\xi \ll 1$), equation (\ref{CONTINUUM})
becomes
\begin{equation}     
\frac{\ve}{\xi}s'(2s-1)+
\frac{\ve^2}{2 \xi^2}s''-k_-s+k_+(1-s)=0;
\label{XIEQN}
\end{equation}          
furthermore, we know that $\ve$ is necessarily on the order
$(k_-+k_+)$.  Since the first order term becomes very small as
$s\rightarrow 0.5$, the second order term must match either the
adsorption or desorption term.  In this case, the second order term
will be balanced when $\xi \sim \sqrt{\ve}$.  Therefore, we expect a
boundary layer of width $O(\sqrt{\ve})$ to arise near the injection
site if $\alpha>0.5$, $s_{\Gamma}<0.5$ or if $\alpha<0.5$ and
$s_{\Gamma}>0.5$.
  
While the boundary layer on the left can be captured using a second
order continuum equation, the boundary layer on the right cannot.  In
equation (\ref{CONTINUUM}), we kept terms only up to order $\ve^2$ in
our expansion $s(x+\ve)$ = $s(x)+\ve
s'(x)+\frac{\ve^2}{2}s''(x)+\frac{\ve^3}{6} s^{(3)}(x)+...$.  The
boundary layer on the right hand side arises to join the outer
solution with the boundary condition $s(L) = 1-w_{+}/w_{-}$.  Making
the substitution $X = x/\xi$, and matching first and second order
terms in (\ref{XIEQN}), we find that $\xi = O(\varepsilon)$.  In
the boundary layer on the left, $s \approx 0.5$, and we assume that
the term $\varepsilon(2s-1)s'$ is relatively small.  When we match the
second order term, $\ve s''/2$, with the adsorption and desorption
terms, $k_{+}(1-s)$ and $k_{-}s$, we find that $\xi \sim
\sqrt{\varepsilon}$.  However, on the right, we cannot assume that $s
\approx 0.5$.  We therefore assume that the leading terms are
$\varepsilon(2s-1)s'$ and $\varepsilon s''/2$, which leads us to
conclude that the wall-hugging boundary layer has width of
$O(\varepsilon)$. In this case, all terms $\ve^n s^{(n)}(X)/n!$ in the
Taylor expansion of (\ref{BCRIGHT}) are $O(1)$, and continuum theory
breaks down.

\bibliography{pasepRefs_PRE}
\end{document}